\documentclass[aps,pre,amsmath,amssymb,reprint,groupedaddress]{revtex4-1}

\usepackage{graphicx}
\usepackage{dcolumn}
\usepackage{bm}

\usepackage{xcolor}

\begin{document}


\title{Dynamical density functional theory for the drying and stratification of binary colloidal dispersions}

\author{Boshen He}
\affiliation{Department of Materials, Loughborough University, Loughborough LE11 3TU, United Kingdom}
\author{Ignacio Martin-Fabiani}%
\affiliation{Department of Materials, Loughborough University, Loughborough LE11 3TU, United Kingdom}%

\author{Roland Roth}
\affiliation{Institut f\"ur Theoretische Physik, Universit\"at T\"ubingen, T\"ubingen, Germany}%

\author{Gyula I. T\'oth}
\affiliation{Department of Mathematical Sciences, Loughborough University, Loughborough LE11 3TU, United Kingdom}
\affiliation{Interdisciplinary Centre for Mathematical Modelling, Loughborough University, Loughborough LE11 3TU, United Kingdom}%

\author{Andrew J. Archer}
\affiliation{Department of Mathematical Sciences, Loughborough University, Loughborough LE11 3TU, United Kingdom}
\affiliation{Interdisciplinary Centre for Mathematical Modelling, Loughborough University, Loughborough LE11 3TU, United Kingdom}%

\date{\today}

\begin{abstract}
We develop a dynamical density functional theory based model for the drying of colloidal films on planar surfaces. We consider mixtures of two different sizes of hard-sphere colloids. Depending on the solvent evaporation rate and the initial concentrations of the two species, we observe varying degrees of stratification in the final dried films. Our model predicts the various structures described in the literature previously from experiments and computer simulations, in particular the small-on-top stratified films. Our model also includes the influence of adsorption of particles to the interfaces.
\end{abstract}


\maketitle

\section{\label{sec:level1}Introduction}

A simple way to form colloidal films is to apply particles suspended in a volatile solvent onto a surface and then to let it evaporate. Such colloidal films are of high relevance to a wide range of applications such as paints \cite{1van2015watching}, inkjet printed structures and patterns \cite{2tekin2008inkjet}, flexible electronics \cite{3ummartyotin2012synthesis}, biomaterials \cite{4procopio2018sol} and adhesives \cite{5jovanovic2004emulsion}. During the film formation process, as the solvent evaporates, the colloidal particles accumulate and are deposited on the substrate to form a close-packed structure  \cite{12fortini2016dynamic}. Typically, mixtures of different  colloids are used, enabling to impart certain properties to the film to meet different functional demands. For example, metal nanoparticles like silver \cite{6suteewong2019pmma} and titania \cite{8evdokimova2018hybrid} are added to enhance the antibacterial properties of coatings. One of the most difficult things to control in these applications is the distribution of the functional nanomaterials within the film, along the direction perpendicular to the substrate. If the functional particles are able to accumulate at the surface during the drying process, then this can lead to an increase in the efficiency of the coating.

There are many factors that influence the final architecture of colloidal films. Some of the most influential are the rate of solvent evaporation, the colloidal diffusion coefficient (rate of Brownian motion), the rate of sedimentation and the form and strength of the interactions between the particles. The competition between these factors determines how the particles are distributed in the final dried structure \cite{10schulz2018critical,13cardinal2010drying, atmuri2012autostratification}. The influence of evaporation and diffusion in the inhomogeneous drying process is usually quantified using the P\'eclet number \cite{routh2001deformation,routh1998horizontal}, which for colloids of type $i$ is defined as the ratio of the evaporation rate and the diffusion rate:
\begin{equation}
\mathrm{Pe}_i=\frac{L_0 v}{D_i},
\label{eq:Pe_number}
\end{equation}
where $v$ is the rate of evaporation (the speed of the air-solvent interface due to evaporation), $L_0$ is the initial thickness of the deposited film and $D_i$ is the diffusion coefficient, which is given by the Stokes-Einstein equation
\begin{equation}
D_i=\frac{k_BT}{6 \pi \eta R_i},
\label{eq:Stokes-Einstein}
\end{equation}
where $\eta$ is the viscosity of the solvent, $R_i$ is the radius of the colloidal particle, and $k_BT$ is the thermal energy, with $k_B$ being Boltzmann's constant and $T$ the temperature. Hence, for $\mathrm{Pe}_i \ll 1$ particles can diffuse away from the receding air-solvent interface, resulting in a homogeneous particle distribution within the film. In contrast, for $\mathrm{Pe}_i \gg 1$ the evaporation is faster than diffusion, so the particles typically become gathered together by the moving interface because there is no time for the particles to diffuse away from it. Considering the influence of sedimentation seems to only be needed when the particles are large, i.e.\ when micron-sized particles are involved \cite{13cardinal2010drying}, or if the mass density of the matter forming the particles is much higher than the mass density of the solvent and evaporation rate is very slow \cite{utgenannt2016fast}. When the colloids have a tendency to aggregate, then the solvent evaporation, which increases the concentration, can increase this and can lead to flocculation during the drying process and therefore to inhomogeneous deposition on the surface. Here, we consider only isotropic particles, treated as mixtures or hard-spheres, where this does not occur. However, the effects of attractive interactions between the colloids can easily be included in the model if required \cite{archer2011interplay, malijevsky2013sedimentation}.

\begin{figure}[t]
	\includegraphics[scale=0.3]{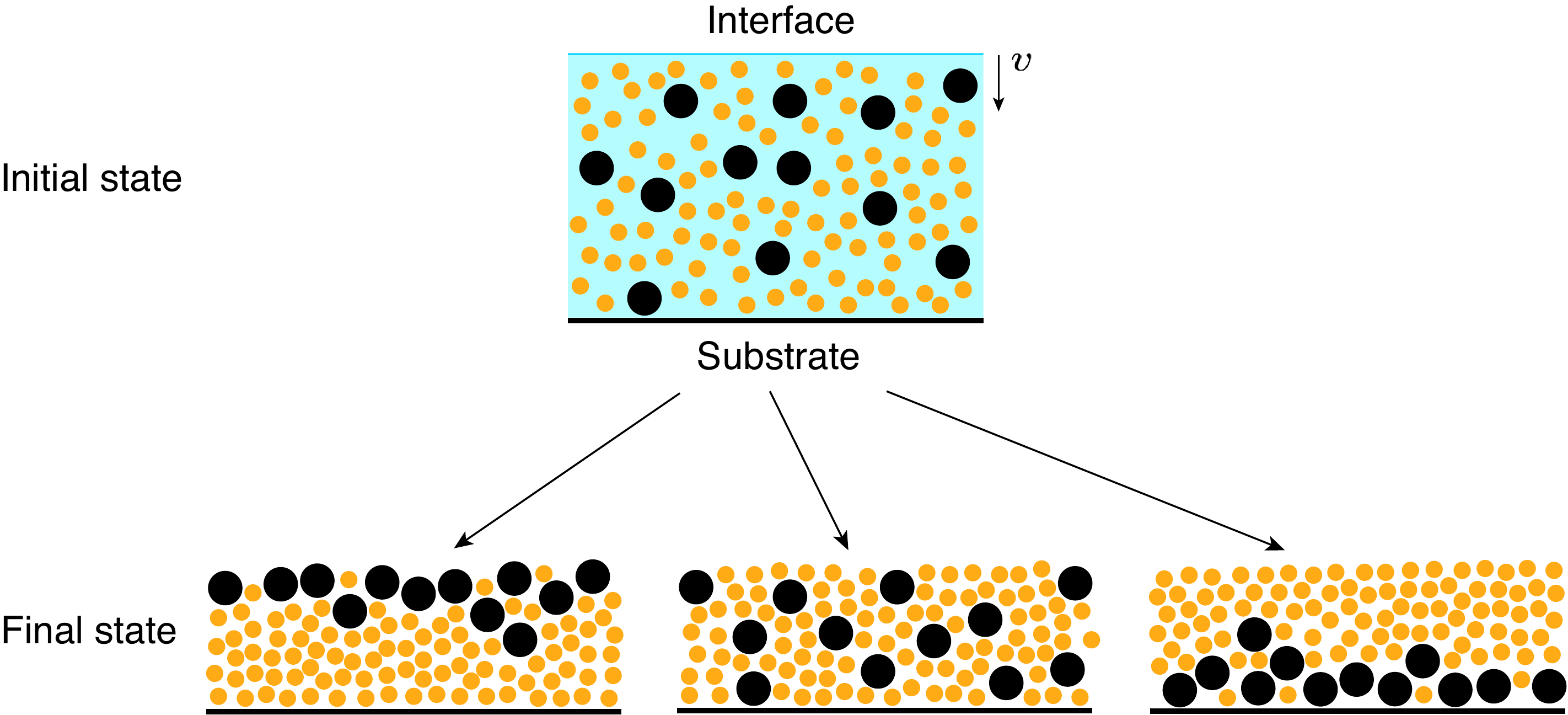}
		\caption{\label{fig:illstration} Schematic overview of the stratification that can occur in the drying of a binary colloidal mixture. The initial colloidal mixture can end up in either the the ``big-on-top'' structure (left), ``small-on-top'' structure (right),  or remain mixed (middle). }		  
\end{figure}

For binary blends composed of two different sizes of particles, henceforth referred to as the `big' ($b$) and the `small' ($s$) particles, respectively, stratification can take place during drying, resulting in either small-on-top or potentially in big-on-top configurations -- see Fig.~\ref{fig:illstration}. 

Trueman et al.~\cite{14trueman2012auto} reported a ``big-on-top'' architecture in colloidal films from such mixtures, referring to this phenomenon as ``auto-stratification'', and argued that it can occur when the two species of particles have different P\'eclet numbers, originating from the different particle diffusion rates. Their results lead to the conclusion that the big-on-top stratification phenomenon can appear when the P\'eclet numbers of the large particles, $\mathrm{Pe}_b$, and the smaller particles, $\mathrm{Pe}_s$, are in the ranges $\mathrm{Pe}_s < 1 < \mathrm{Pe}_b$. These considerations do not take into account the influence of the interactions between particles and packing effects. When the suspensions become dense enough that these effects become important, the experiments and Langevin dynamics simulations of Fortini et al.~\cite{12fortini2016dynamic} show that an alternative stratification phenomenon can occur when $1 \ll  \mathrm{Pe}_s \ll \mathrm{Pe}_b$, where the smaller particles accumulate on top, near the interface. 

Fortini et al.~\cite{12fortini2016dynamic} showed that gradients in the osmotic pressure, originating from non-uniform particle concentrations, push the particles differently based on their size, with the larger particles moving faster towards the bottom of the dispersion (nearer to the substrate surface). Zhou et al.~\cite{15zhou2017cross} developed a simple model for this, using a second-virial coefficient level approximation for the free energy, applicable for describing the influence of the interactions in dilute and moderately dilute binary colloidal system. Their model can be viewed as a simple dynamic density functional theory (DDFT) \cite{18marconi1999dynamic,20marconi2000dynamic, 19archer2004dynamical, archer2004dynamical, hansen2013theory} together with a local density approximation (LDA) for the Helmholtz free energy functional \cite{hansen2013theory, evans1979nature, evans92, wu2006density}. We say much more about DDFT below. The results of Zhou et al.~\cite{15zhou2017cross} show that at high enough values of the P\'eclet numbers and small particle volume fractions, the cross-interactions exert forces on the larger particles that are strong enough to push them towards the substrate during drying, away from the air-solvent interface, resulting in small-on-top configurations. 

In Howard et al.~\cite{9howard2017stratification}, molecular dynamics computer simulation results are reported that confirm the occurrence of small-on-top stratification and also that a higher size ratio can lead to more of the small particles segregating at the top, near the air-solvent interface. They use a DDFT based argument to determine the difference in the migration velocities of the two different sizes of particles in the drying film. The differing velocities lead to the stratification. The results in Refs.~\cite{12fortini2016dynamic, 9howard2017stratification} were obtained by treating solvent implicitly. In contrast, Tang et al.~\cite{16tang2018stratification} developed a molecular dynamics simulation model with the solvent treated explicitly. Their results show the small-on-top stratification still occurs, although the regime in which this happens is $\mathrm{Pe}_s\varphi_{s} \gtrsim 1$, where $\varphi_{s}$ is the volume fraction of the smaller particles, as predicted by Sear and Warren when considering solvent backflow.\cite{sear2017diffusiophoresis, sear2018stratification} This threshold is significantly lower in models with an implicit solvent, such as that of Zhou et al.~\cite{15zhou2017cross}, where the condition for size segregation is $\mathrm{Pe}_s\varphi_{s} \gtrsim \alpha^{-2}$, where $\alpha\equiv R_b/R_s$ is the colloid size ratio and $R_b$ and $R_s$ are the radii of the big and small colloids, respectively. Thus, treating solvent explicitly or implicitly does not change the direction of the predicted stratification but it can influence the degree to which it occurs. It should also be mentioned that Tang et al.~\cite{16tang2018stratification} find that treating the solvent evaporation explicitly can sometimes enhance the stratification, due to colloidal thermophoresis associated with the evaporative cooling. Even more recently, Liu et al.~\cite{17liu2019sandwich} described two new stratification phenomena: big-small-big and small-big-small sandwich film structures. The mechanism(s) behind this sandwich structure formation is still not clear.

In this paper, we use DDFT \cite{18marconi1999dynamic, 20marconi2000dynamic, 19archer2004dynamical, archer2004dynamical, hansen2013theory} to investigate the time evolution of mixtures of two different sizes of colloidal particles suspended in a liquid film spread with uniform thickness onto a planar surface. We include in our model the evaporation of the films and so in the relevant parameter regimes we observe stratified drying of the colloids. DDFT is a theory for the time evolution of the average particle density distributions in systems of interacting Brownian particles, i.e.\ it is a theory for colloids suspended in a solvent, with the solvent treated implicitly. DDFT builds on equilibrium density functional theory (DFT) \cite{hansen2013theory, evans1979nature, evans92, wu2006density}, and so when a reliable approximation for the Helmholtz free energy functional is used (as we do here) then it can accurately predict the density profiles of the colloids, at least when the system is at equilibrium. Moreover, based on past experience \cite{royall2007nonequilibrium}, we can also be confident that it is also at least qualitatively accurate for the out-of-equilibrium situations considered here. 
The highly accurate fundamental measure theory (FMT) for mixtures \cite{roth2010fundamental} (see also Refs.~\cite{roth2002fundamental, hansen2006density}) is used here to approximate the Helmholtz free energy functional. In Ref.~\cite{royall2007nonequilibrium}, DDFT together with FMT was used to describe the dynamics of colloids sedimenting under gravity within a liquid film. This work showed that in circumstances where the colloids are initially uniformly distributed within the liquid and solvent hydrodynamic flows are weak, then DDFT is very reliable. However, in cases where the hydrodynamic flows become sizeable, then the hydrodynamic interactions between the colloids which are not present in standard DDFT need to be incorporated \cite{sear2017diffusiophoresis}. These can be included \cite{rex2008dynamical, goddard2012general, goddard2016dynamical}, but we do not do so here, since our results are largely in regimes where the suspension are fairly dense and we expect the hydrodynamic interactions to largely be screened.

We assume that the variations in the average density profiles only occur in the direction perpendicular to the substrate and so we are able to treat the system as being effectively one dimensional. We incorporate the influence on the colloids of the moving air-solvent interface via an external potential. Thus, the colloids are treated as being effectively confined between two parallel planar walls: the lower one, which is stationary, corresponding to the surface of the substrate and an upper one, corresponding to the air-solvent interface, which is treated as moving with constant velocity $v$ (i.e.\ constant evaporation rate) towards the substrate (lower wall). We also assume that the interactions between these two interfaces and the colloids is purely repulsive. Thus, we assume that the colloids are strongly solvophilic, so they remain in the solvent and do not cross the air-solvent interface. The effective potential between such colloidal particles and a liquid-liquid interface was calculated using DFT in Ref.~\cite{hopkins2009solvent} and here we assume that the external potential between the colloids and the air-solvent interface is similar.

We initiate the system with the two walls a distance $L_0$ apart, with the initial colloidal density profiles being those of the system at equilibrium. Thus, we assume that the colloids are uniformly distributed in the bulk of the fluid between the walls, corresponding to a homogenous suspension of colloids. We also neglect the influence of gravity, which is irrelevant for the few hundred nanometer size of colloids we consider. We then calculate how the colloidal density profiles evolve over time as the system is compressed between the two interfaces. However, this compression can not proceed indefinitely; eventually the system becomes sufficiently close-packed that in at least portions of the system either freezing or jamming occurs. Shortly, before the system reaches this situation, the algorithm that we use to integrate forward in time the DDFT equations fails and we define that time as the end point time $t_{final}$ and assume that the density profiles at this time are the same as those of the final dried film. This is a reasonable assumption, since the mobility of the colloids in such densely packed films is low and only small local particle rearrangements can be expected after this stage.

For most of the results presented here, we set the size ratio of the radii of the two species of colloids $\alpha= R_b/R_s=2$, however we do also present results for the case $\alpha=5$. We observe the small-on-top stratification that occurs when the solvent evaporation rate is fast enough, that was observed in the simulations and experiments of Refs.~\cite{tatsumi2018effects, 12fortini2016dynamic, 9howard2017stratification}. Also, as a transient in initially dilute systems, we see states that are similar to the inverted big-on-top stratification that occurs when $\mathrm{Pe}_s < 1 < \mathrm{Pe}_b$. However, these do not persist into the final dense state. We also show how the stratification depends on the initial concentrations of the two colloidal species. The advantage of using DDFT is that in addition to the colloidal density profiles, we also have access to thermodynamic quantities, including the gradient of the local chemical potentials, which are the thermodynamic forces driving the fluxes in the drying film.


\section{\label{sec:level2}DDFT model for the system}
DFT shows that the thermodynamic grand potential functional for a mixture can be written as \cite{hansen2013theory, evans1979nature, evans92, wu2006density} 
\begin{equation}
\Omega\left[\{\rho_i\}\right]=\mathcal{F}\left[\{\rho_i\}\right]+\sum_{i} \int d^3r\rho_{i}(\boldsymbol{r})\left(V_{ext}^{i}(\boldsymbol{r})-\mu_{i}\right),
\label{grand_potential}
\end{equation}
where $\{\rho_i\}=\{\rho_s,\rho_b\}$ are the density profiles of the two different species of particles in the system and where the index $i=s,b$ labels the two different species of small and big particles, respectively. $\mathcal{F}[\{\rho_i\}]$ is the intrinsic Helmholtz free energy functional for the mixture, $\mu_{i}$ are the chemical potentials and $V_{ext}^{i}(\boldsymbol{r})$ are the effective external potentials acting on each of the different species, which includes both the influence of the solid substrate on which the colloidal suspension is deposited and also the influence of the descending air-solvent interface.

The intrinsic Helmholtz free energy functional can be separated into two parts,
\begin{equation}
\mathcal{F}\left[\{\rho_i\}\right]=\mathcal{F}_{id}\left[\{\rho_i\}\right]+\mathcal{F}_{ex}\left[\{\rho_i\}\right],
\label{equ3}
\end{equation}
where
\begin{equation}
\mathcal{F}_{id}[\{\rho_i\}]=k_BT\sum_{i} \int d^3r\rho_{i}(\boldsymbol{r})\left(\ln[\Lambda_i^3\rho_i(\boldsymbol{r})]-1\right)
\end{equation}
is the intrinsic Helmholtz free energy of an ideal-gas (i.e.\ for a system of non-interacting particles), where $\Lambda_i$ is the thermal de Broglie wavelength for particles of species $i$ and $\mathcal{F}_{ex}[\{\rho_i\}]$ is the excess contribution due to the interactions between the particles. Here, we assume that the colloids interact via hard-sphere pair potentials and treat the system using the highly accurate FMT approximation for $\mathcal{F}_{ex}[\{\rho_i\}]$, which is comprehensively reviewed in Ref.~\cite{roth2010fundamental}. At equilibrium, the fluid density profiles are those which minimise the grand potential in Eq.~\eqref{grand_potential}, i.e\ which satisfy the coupled pair of Euler-Lagrange equations
\begin{equation}\label{eq:func_deriv}
\frac{\delta \Omega[\{\rho_i\}]}{\delta \rho_{i} (\boldsymbol{r})}=\frac{\delta F[\{\rho_i\}]}{\delta \rho_{i} (\boldsymbol{r})}-\mu_i=0.
\end{equation}
The central task in equilibrium DFT is to solve these equations. 

However, to describe the dynamics of a non-equilibrium inhomogeneous colloidal fluid, DDFT is required \cite{18marconi1999dynamic, 20marconi2000dynamic, 19archer2004dynamical, archer2004dynamical, hansen2013theory}. It is a theory for the time dependent one-body density profiles $\rho_i(\boldsymbol{r},t)$ which are defined via an ensemble average over all possible configurations at time $t$, given a specific distribution of states at an earlier time. Since the average density is a conserved quantity, the dynamics is given by the continuity equation
\begin{equation}
\frac{\partial \rho_{i} (\boldsymbol{r},t)}{\partial t} =-\nabla \cdot \boldsymbol{J}_{i} (\boldsymbol{r},t),
\label{equ4}
\end{equation}
where $\boldsymbol{J}_{i}$ is the number current for species $i$. The approximations made in deriving DDFT \cite{18marconi1999dynamic, 20marconi2000dynamic, 19archer2004dynamical, archer2004dynamical, hansen2013theory} are equivalent to setting the currents $\boldsymbol{J}_{i}$ to be proportional to the gradient of the local non-equilibrium chemical potentials
\begin{equation}
\boldsymbol{J}_{i} (\boldsymbol{r},t)=-\Gamma_{i}\rho_{i}(\boldsymbol{r},t) \nabla \mu_{i}(\boldsymbol{r},t).
\label{equ5}
\end{equation}
where $\Gamma_{i}=D_i/k_BT$ is a mobility constant for species $i$ and the non-equilibrium chemical potentials (within the DDFT approximation) are given by Eq.~\eqref{eq:func_deriv}. In other words, it is gradients in $\mu_{i}=\delta F/\delta\rho_i$ that are the thermodynamic driving forces for the diffusion of the colloids. The DDFT assumption that Eq.~\eqref{eq:func_deriv} still holds out of equilibrium, i.e.\ that there exist non-equilibrium chemical potentials and that they are given by the functional derivative of the Helmholtz free energy functional with respect to the respective density profile, constitutes an adiabatic approximation equivalent to assuming that the two-particle correlations in the non-equilibrium fluid are the same as those in an equilibrium system having the same one-body density profiles $\rho_i(\boldsymbol{r})$ \cite{18marconi1999dynamic, 20marconi2000dynamic, 19archer2004dynamical, archer2004dynamical, hansen2013theory}. Thus, the time evolution of the density profiles is given by the following coupled pair of equations
\begin{equation}
\frac{\partial \rho_{i} (\boldsymbol{r},t)}{\partial t}=\Gamma_{i} \nabla \cdot \left[\rho_{i} (\boldsymbol{r},t) \nabla \frac{\delta F[\{\rho_i\}]}{\delta \rho_{i} (\boldsymbol{r},t)}\right],
\label{equ6}
\end{equation}
which are the central equations in DDFT. Note that these equations assume that the colloids have the following over-damped stochastic equations of motion:
\begin{equation}
\frac{\dot{\boldsymbol{r}}_{j}}{\Gamma_{i}}= -\nabla U(\{\boldsymbol{r}_j\})+\mathbf{R}_{t},
\end{equation}
where $\dot{\boldsymbol{r}}_{j}(t)$ is the velocity of particle $j$, $U(\{\boldsymbol{r}_j\})$ is the potential energy of the system, which includes both the contributions from the external potentials and also the contributions from the interactions between the particles. $\mathbf{R}_{t}=(\xi_j^x(t),\xi_j^y(t),\xi_j^z(t))$ is a delta-correlated stationary Gaussian process with zero-mean $\langle\xi_j^p(t)\rangle=0$ and correlator $\langle\xi_j^p(t)\xi_j^q(t')\rangle=2k_BT\delta_{pq}\delta(t-t')$, that describes the influence of the random thermal fluctuations forces on the colloids due to the solvent \cite{18marconi1999dynamic, 20marconi2000dynamic, 19archer2004dynamical, archer2004dynamical, hansen2013theory}.

\begin{figure}[t]
	\includegraphics[scale=0.7]{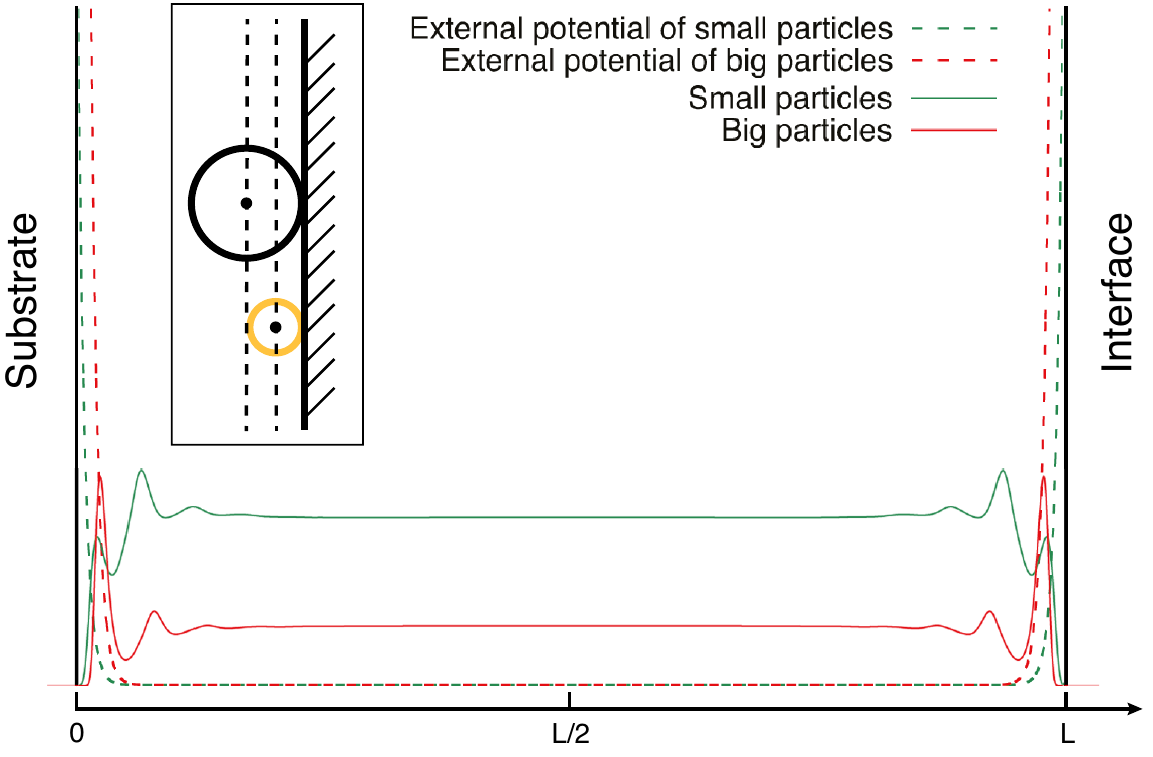}
	\caption{\label{fig:external}
Illustration of the external potential and the corresponding equilibrium density profiles at time $t=0$, for the case when the bulk fluid volume fractions of the small and big particles are $\varphi_{s}=0.1$ and $\varphi_{b}=0.3$, respectively. The substrate and the air-solvent interface are both treated as planar `walls', with external potential given in Eq.~\eqref{equ7}. The inset is a schematic diagram showing how particles arrange at the wall, with an off-set in the distances at which the centres of the particles `feel' the wall. }		
\end{figure}

As mentioned in the introduction, we treat both the influence of the moving air-solvent interface and the substrate interface via effective external potentials. We assume the following form for these potentials:
\begin{equation}
\beta V_{ext}^i(z,t)=Be^{-\kappa(z-R_i)}+Be^{\kappa(z+R_i-L(t))}+m_igz,
\label{equ7}
\end{equation}
where $\beta=(k_BT)^{-1}$, $z$ is the cartesian direction perpendicular to the interfaces, $L(t)$ is the distance between the interfaces and $B$ and $\kappa$ are two parameters that determine the strength and softness of the external potentials. Here, we use the values $B=7400$ and $\kappa=2R_s^{-1}$. In particular, the value of $\kappa$ controls the softness of the interfaces. Since we are primarily interested in the ordering near the moving air-solvent interface, for simplicity we make the potentials due to the two interfaces of the same form. We assume that the top interface is moving downwards with a constant velocity, as an alternative to considering explicitly the mass flow through such interface. This approach, known as the one-sided limit approximation,\cite{burelbach1988nonlinear} does not take into account the gas phase dynamics except through the boundary condition at the interface. The one-sided limit approximation is used extensively in the literature on the drying colloidal blends.\cite{12fortini2016dynamic, 9howard2017stratification, tatsumi2018effects} One could extend our model, by incorporating more of the physics of evaporation, e.g.\ by treating the effects of diffusion in the vapor\cite{sultan2004evaporation} or by using the Hertz-Knudsen relation\cite{persad2016expressions} to relate the evaporation rate to the environmental conditions. Thus, we assume that the distance between the two interfaces depends on time as follows
\begin{equation}
L(t)=L_0-vt,
\label{eq:L_t}
\end{equation}
where $v$ is the constant velocity of the air-solvent interface, which is moving due to the solvent constant evaporation rate. We typically choose the initial distance between the interfaces to be either $L_0=77R_s$ or $L_0=154R_s$. The final term in Eq.~\eqref{equ7} allows us to incorporate the influence of gravity, with $m_i$ being the buoyant mass of the colloids of species $i$ and $g$ is the acceleration due to gravity. However, since our interest here is in colloidal systems where the particles are small enough that these contributions are negligibly small (i.e.\ with diameters of only a few hundred nanometers), we therefore set $m_i=0$ throughout.

To solve the DDFT equations \eqref{equ6}, the particle densities $\rho_s(\boldsymbol{r},t)$ and $\rho_b(\boldsymbol{r},t)$ for the binary mixture are initialised at $t=0$ with the density profiles of a system at equilibrium in the external potentials given by Eq.~\eqref{equ7} at time $t=0$, to model a well-mixed suspension having been deposited on the surface. A similar approach was also used in Ref.~\cite{royall2007nonequilibrium}. The equilibrium density profiles are obtained by initiating the system with the density profiles of an ideal-gas and then evolving them in time for a short period till the system is equilibrated. Note however that as long as the initial density profiles correspond to a (roughly) uniform distribution of the colloids between the interfaces, the precise form of the initial density profiles have negligible influence on the subsequent dynamics. To evolve the DDFT equations forward in time we use the Crank-Nicholson algorithm \cite{press1992numerical} with a typical value for the time step of $10^{-4}\tau$, where $\tau = \beta R_{s}^{2}/\Gamma_s$ is the Brownian time unit for the small particles. We use fast Fourier transforms to evaluate the convolutions arising from $\mathcal{F}_{ex}[\{\rho_i\}]$, as described in detail in Ref.~\cite{roth2010fundamental}.

In Fig.~\ref{fig:external} we illustrate the external potential and also the corresponding initial (equilibrium fluid) density profiles with which we initiate the system. These profiles are for the particular case when the initial bulk fluid average packing fractions of the two species are $\varphi_{b}=0.1$ and $\varphi_{b}=0.3$, respectively. The inset diagram in Fig.~\ref{fig:external} of two different sized particles at a wall illustrates why the particle radii appear in the external potentials in Eq.~\eqref{equ7}: if there is a `wall' to the right of the particles at $z=L$, then the centres of the particles will be at $z=L-R_i$ when they encounter the wall. Treating the interaction between the colloids and the air-solvent interface in this way assumes that the colloids are strongly solvophilic. If the colloids are not so solvophilic, then the effective potential between the colloids and the interface can be attractive \cite{hopkins2009solvent} and self assembly and other ordering of the colloids at the interface can occur \cite{bresme2007nanoparticles}. As it stands, the model we use here does not include these effects. However, it would be straight-forward to include the effects of particle adsorption at the interface in the model, if required.

The two density profiles in Fig.~\ref{fig:external} are typical of a moderately dense system. Both density profiles exhibit a series of peaks in the vicinity of the interfaces. The oscillations in the profiles are a signature of the effects of packing of the particles at the walls and the fact that the profiles take constant values in the bulk between the interfaces indicates the particles are on average uniformly distributed in this region \cite{hansen2013theory, evans92}. The presence of a peak in both density profiles right at the interfaces indicates that there are particles of both species in contact with them. However, since the peak in the density profile of the large particles is far more prominent, this indicates there is a preference for the larger particles to be right at the interfaces. This is a consequence of the depletion interaction (an effective attraction between the interface and the big particles, due to excluded volume effects stemming from the presence of the smaller particles\cite{asakura1954interaction, gotzelmann1999depletion, roth2000depletion}) which is then enhanced by the repulsive tails of the interface potentials, since these also leads to an effective attraction between the interface and the big particles, due to the fact that we use the same values of $B$ and $\kappa$ for both the $b$ and $s$ potentials in Eq.~\eqref{equ7}.\cite{roth2000binary, archer2002wetting}

\section{\label{sec:3}Results}
\subsection{Dense systems}
\subsubsection{Influence of evaporation rate}

\begin{figure*}[t!]
	\includegraphics[scale=0.6]{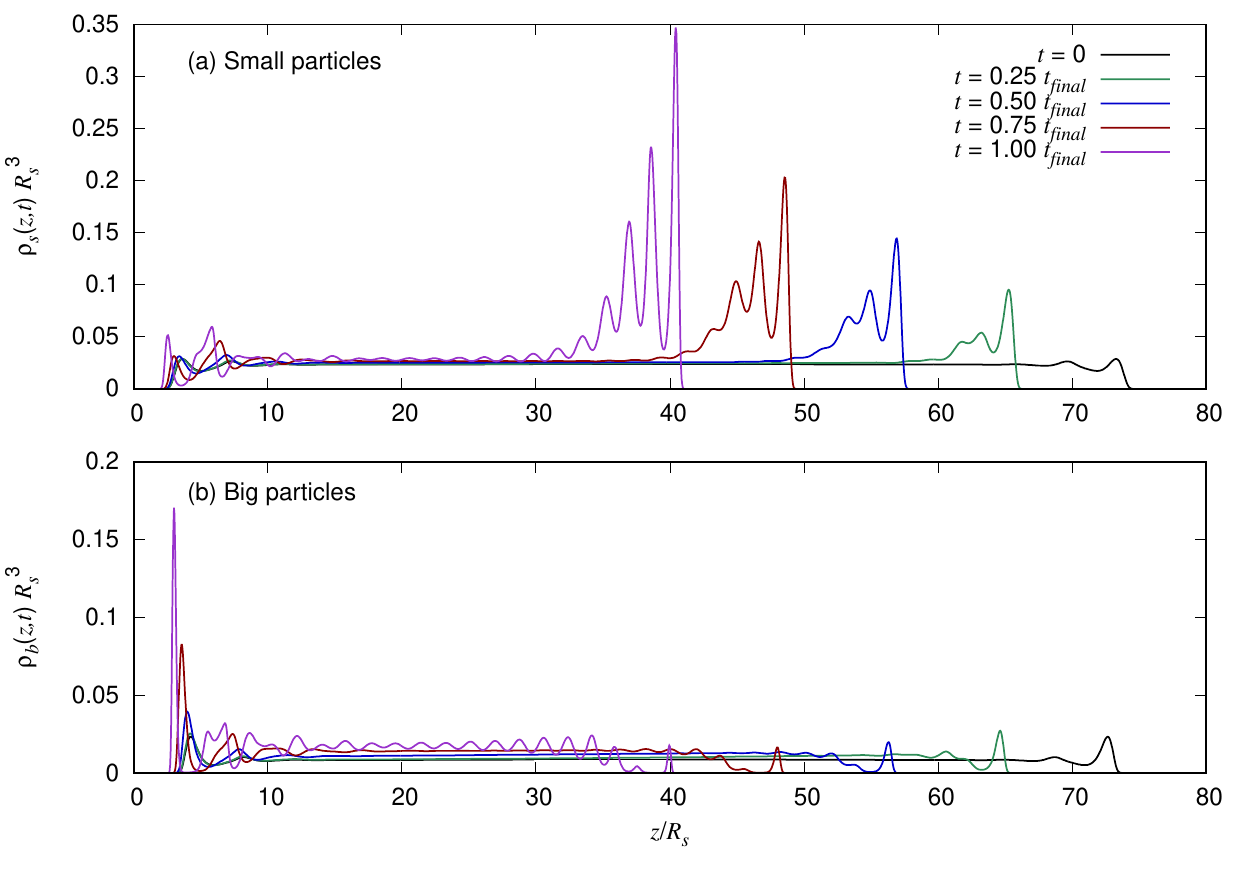}
	\includegraphics[scale=0.6]{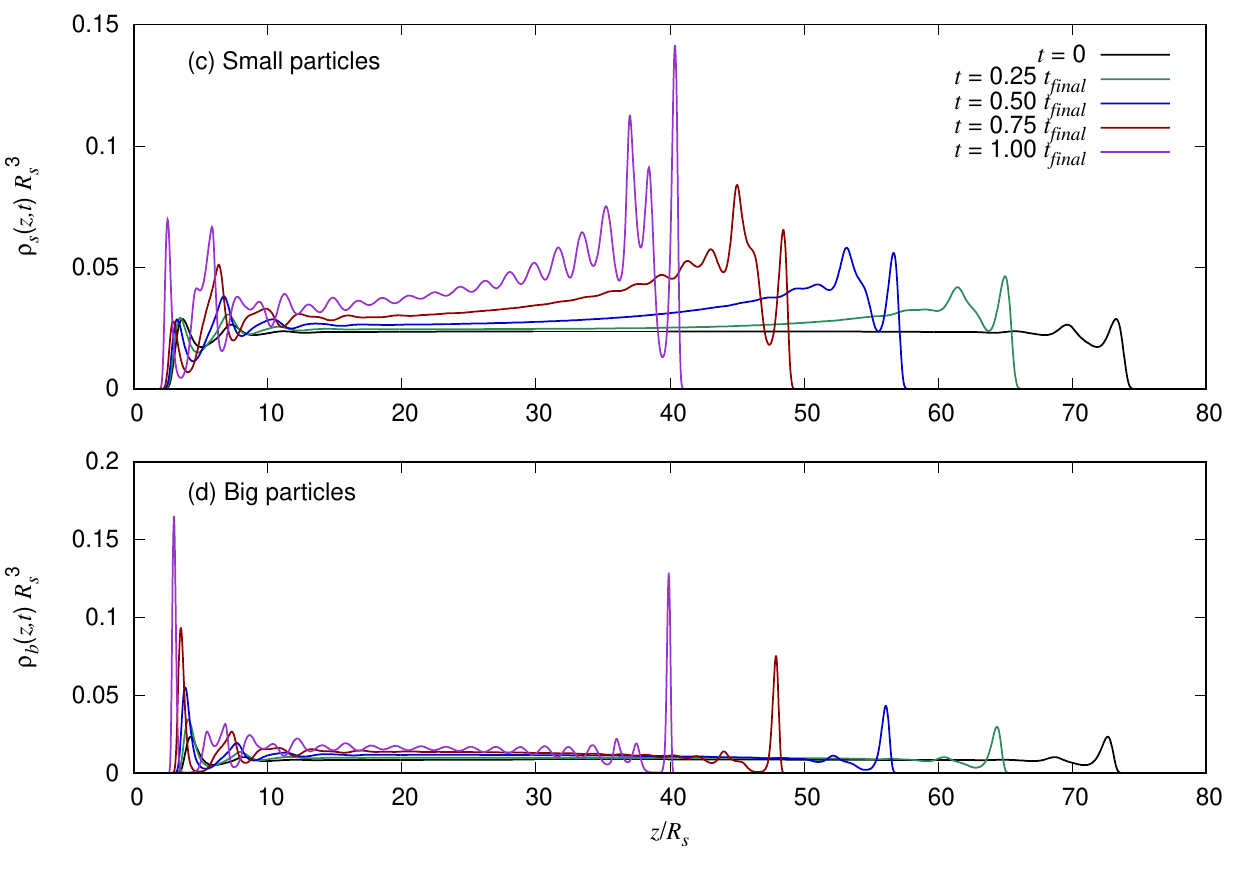}	
	\caption{\label{fig:different_speed} Time sequence of density profiles during the evaporation of systems with initial volume fractions $\varphi_s = 0.1$ and $\varphi_b = 0.3$, particle size ratio $\alpha=2$ and initial film thickness $L_0=77R_s$. Panels (a) and (b) on the left show the profiles for the small and big particles respectively, for the case when the evaporation rate (interface descending velocity) is $v=0.5R_s/\tau$, corresponding to $\mathrm{Pe}_s\approx40$ and $\mathrm{Pe}_b\approx80$, with the final time $t_{final}=68\tau$. Panels (c) and (d) on the right are for the case when $v=0.05R_s/\tau$, corresponding to $\mathrm{Pe}_s\approx4$ and $\mathrm{Pe}_b\approx8$, with $t_{final}=680\tau$. Some of the highest peaks in density profiles are not displayed in order to show more details of the profiles away from the peaks.}  	
\end{figure*}

\begin{figure}[t!]
	\includegraphics[scale=0.6 ]{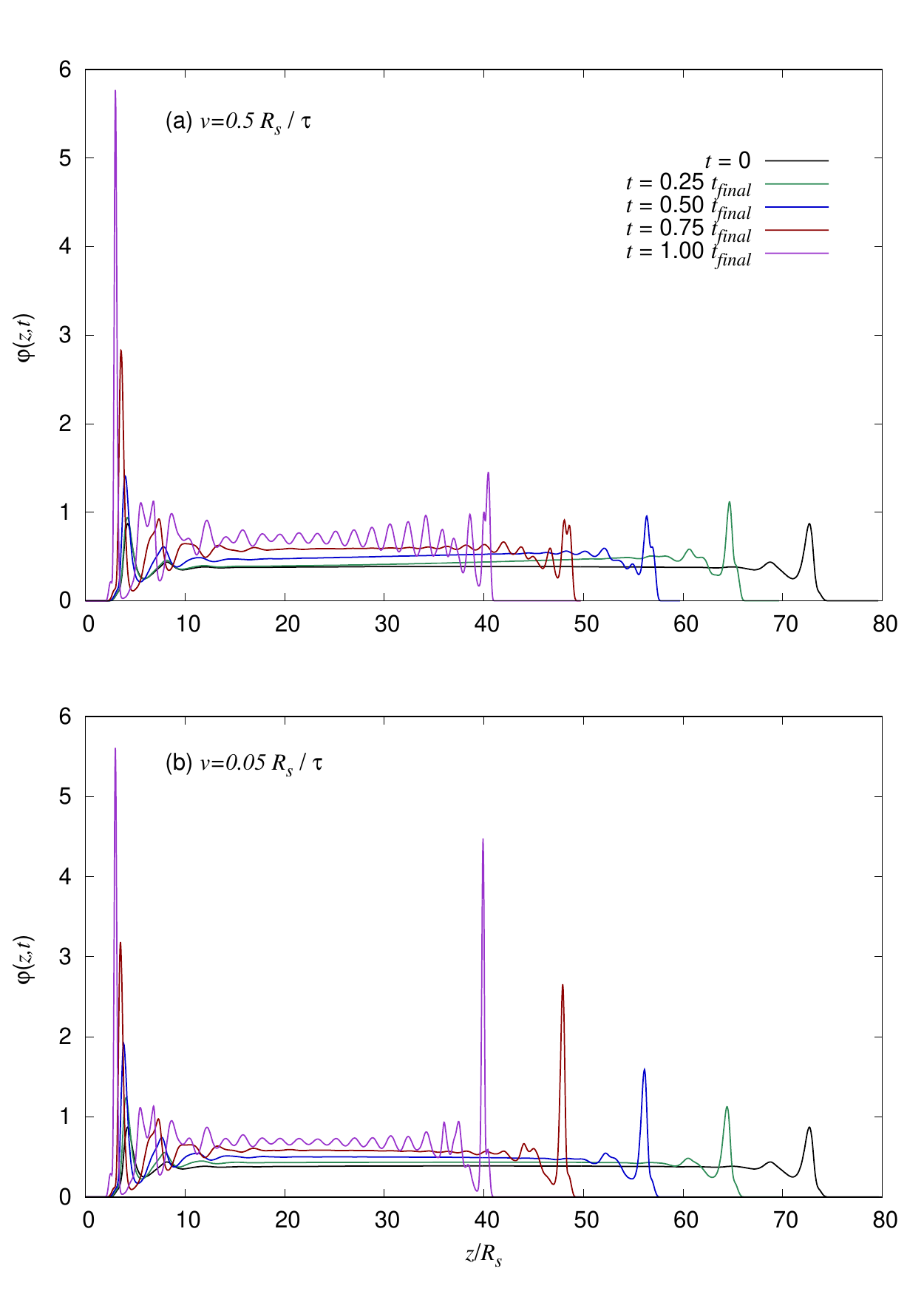}	
	\caption{\label{fig:total_packing} Time evolution of the local total packing fraction, corresponding to the two different velocity $v$ cases presented in Fig.~\ref{fig:different_speed}.}		
\end{figure}

\begin{figure}[t]
	\includegraphics[scale=0.95]{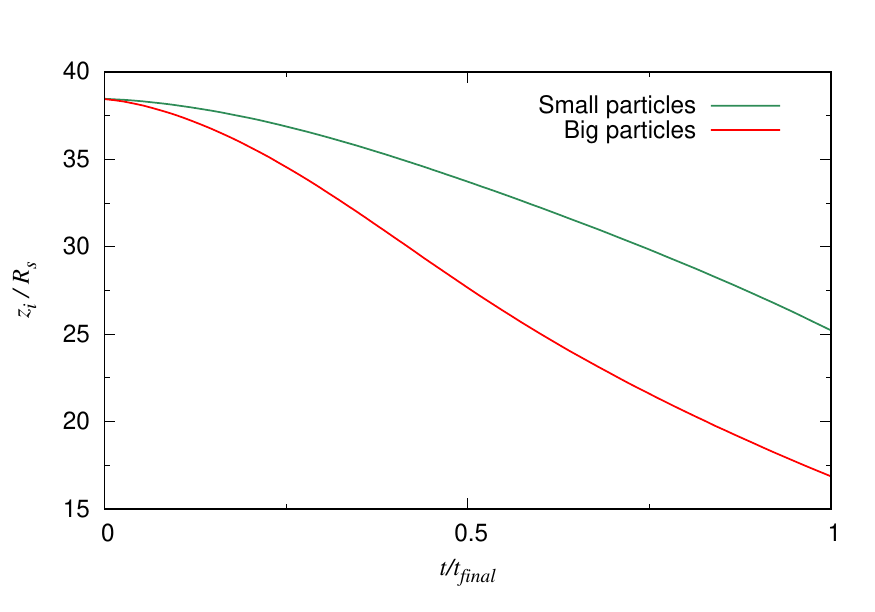}	
	\caption{\label{fig:center_of_mass} The time evolution of the centre of mass of the two different species of particles, corresponding to the case where the interface descending velocity is $v=0.5R_s/\tau$, with density profiles displayed in Fig.~\ref{fig:different_speed}(a)--(b).}
\end{figure}

\begin{figure}[t]
	\includegraphics[scale=0.95]{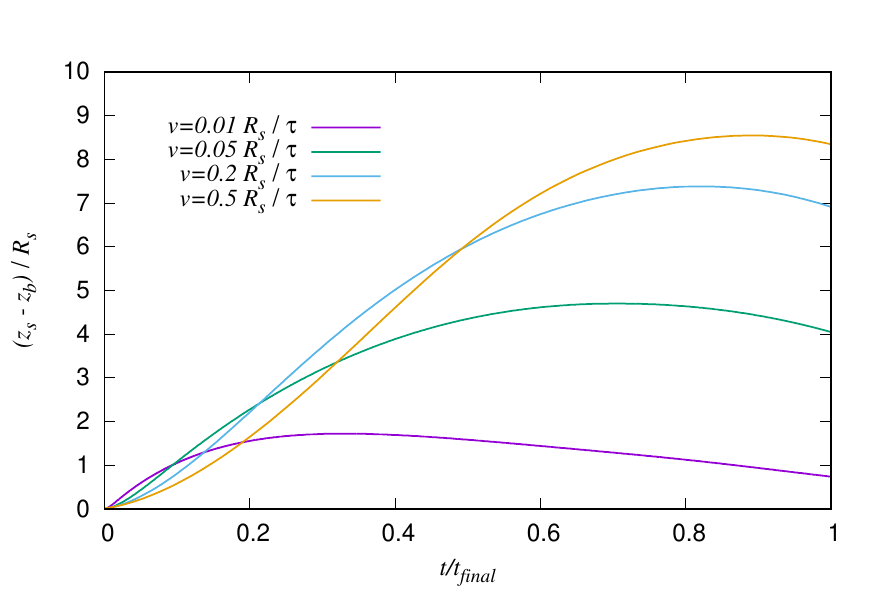}
	\caption{\label{fig:diff_centre_of_mass} The distance between the centres of mass of the different species $(z_s-z_b)$ over time, for systems with various different interface descending velocities, as given in the key. The size ratio of the particles is $\alpha=2$ and the initial average volume fractions are identical in all cases, $\varphi_s = 0.1$ and $\varphi_b = 0.3$. For increasing $v$ as given in the key, which correspond to $\mathrm{Pe}_s\approx0.8$, 4, 16 and 40 and $\mathrm{Pe}_b\approx1.6$, 8, 32 and 40. The final times are $t_{final}/\tau=3583$, $717$, $180$ and $72$.6}	
\end{figure}

In order to illustrate the influence of the evaporation rate on the final film structure, we display in Fig.~\ref{fig:different_speed} the time evolution of the density profiles for two different values of the evaporation rate (interface velocity) $v$. These results are for a fairly dense system with initial particle volume fractions of $\varphi_{s}=0.1$, $\varphi_{b}=0.3$ and size ratio $\alpha=2$. In panels (a) and (b) are the density profiles of the big and the small particles, respectively, for the case when $v = 0.5R_s/\tau$. This speed roughly equates to an evaporation rate $\approx10^{-5}$m/s for colloids of radius $R_s=100$nm or, since $L_0=77R_s$, from Eq.~\eqref{eq:Pe_number} it also corresponds to the P\'eclet numbers $\mathrm{Pe}_s\approx40$ and $\mathrm{Pe}_b\approx80$. In the initial state, the particle density profiles are uniform in the bulk of the liquid away from the interfaces, with some oscillations due to packing effects in the vicinity of the substrate and interface. As time progresses, the small particles start to accumulate at the descending upper interface while the big particles are depleted from it. This observation is in line with the results of Fortini et al.~\cite{12fortini2016dynamic}, where they observed size segregation of binary colloidal blends as a result of diffusiophoresis.  Note that we observe a single peak in the density profile of the big particles right at the interface, remaining in that position throughout the whole drying process. This occurs because a few of the big particles adsorb to the interface due to the depletion attraction and also the fact that we set the value of $\kappa$ the same in the repulsive interface potentials for both species, which leads to a further effective attraction between the interface and the big particles. A layer at the interface mainly made of the big particles, remaining trapped throughout the whole drying process, was also observed in \cite{9howard2017stratification}. Here, the top layer is composed of both big and small particles. The moment when the simulation stops corresponds to the physical limit where the particles approach the highest total packing fraction possible -- see Fig.~\ref{fig:total_packing}, where we display the time evolution of the local total packing fraction $\varphi(z)=\rho_s(z)\frac{4}{3}\pi R_s^3+\rho_b(z)\frac{4}{3}\pi R_b^3$. This time is noted as $t_{final}$, and it is defined in the same way for all the simulations presented here.

In panels (c) and (d) of Fig.~\ref{fig:different_speed} we show the time evolution of the density profiles for a system with a slower interface descending speed, $v = 0.05R_s/\tau$, compared to the case in (a) and (b). This corresponds to the P\'eclet numbers $\mathrm{Pe}_s\approx4$ and $\mathrm{Pe}_b\approx8$ and roughly equates to an evaporation rate $\approx10^{-6}m/s$ for colloids of radius $R_s=100$nm, similar to the evaporation rate of water close to an infrared lamp \cite{utgenannt2016fast}. There is still some accumulation of small particles close to the interface and a depletion of big particles from that layer. However, the effect is much less marked than in the faster evaporation case displayed in panels (a) and (b). Note that with the interface descending more slowly, the system is not driven as far from equilibrium as in the case with faster evaporation, resulting in a less efficient size segregation. This observation is in line with the work of Tatsumi et al.\ who reported the existence of optimal P\'eclet numbers above and below which the size segregation is not as efficient \cite{tatsumi2018effects}. Note also that Fig.~\ref{fig:different_speed}(d) shows that the number of big particles right at the interfaces increases as time progresses. This build-up is due to the fact that the total density is increasing over time and so the depletion interaction between the big particles and the interface also over time increases in strength \cite{roth2000binary, roth2000depletion}.

The plots of the time evolution of the total packing fraction in Fig.~\ref{fig:total_packing}, corresponding to the results for two different interface velocities in Fig.~\ref{fig:different_speed}, shows as one would expect that the overall average packing fraction between the interfaces increases with time, with pronounced peaks at both interfaces. The peak at the upper interface is larger and narrower for the slower interface velocity case. In contrast, in the faster system the peak at the upper interface is broader, with a double-peaked structure at later times.

Quantitative insight into the nature of the time evolution of the particles can be gained from determining the position of the centres of mass
\begin{equation}
z_i(t)=\frac{\int_0^{L_0} z \rho_i(z)dz}{\int_0^{L_0} \rho_i(z)dz}
\end{equation}
of the two different species during the drying process. Figure~\ref{fig:center_of_mass} shows these, for the system with $v = 0.5R_s/\tau$ and corresponding density profiles in Fig.~\ref{fig:different_speed}(a)--(b). We see that at $t = 0$, before the interface starts moving, the centres of mass of both the small and big particles are at the same distance from the lower surface, because the particles are at equilibrium and uniformly distributed between the two interfaces. As the upper interface descends, of course the positions of both centres of mass decrease over time, as the particles are pushed towards the substrate. Moreover, due to particle interactions there is segregation and so we see that the centre of mass distance for the big particles $z_b$ decreases at a much faster rate than the distance $z_s$, for the small particles. In Fig.~\ref{fig:diff_centre_of_mass} we display the difference $(z_s-z_b)$ over time. In the limit $v\to0$, where the system remains in quasi-equilibrium throughout the drying process, then this difference remains equal to zero for all times $t$. The fact that we see $(z_s-z_b)>0$ for $v>0$ is a signature of the system being out-of-equilibrium and the larger the value of this difference, the further from equilibrium the system is and the greater the segregation effect. Figure~\ref{fig:diff_centre_of_mass} shows results for four different values of the interface velocity ($v\tau/R_s=0.01$, $0.05$, $0.2$ and $0.5$). We see that as the velocity increases the segregation effect increases. However, since the curves for the two higher velocity values are somewhat similar, this indicates that the observed segregation effect saturates somewhat for $v\tau/R_s\gtrsim0.1$.

\begin{figure}[t]
	\includegraphics[scale=0.73]{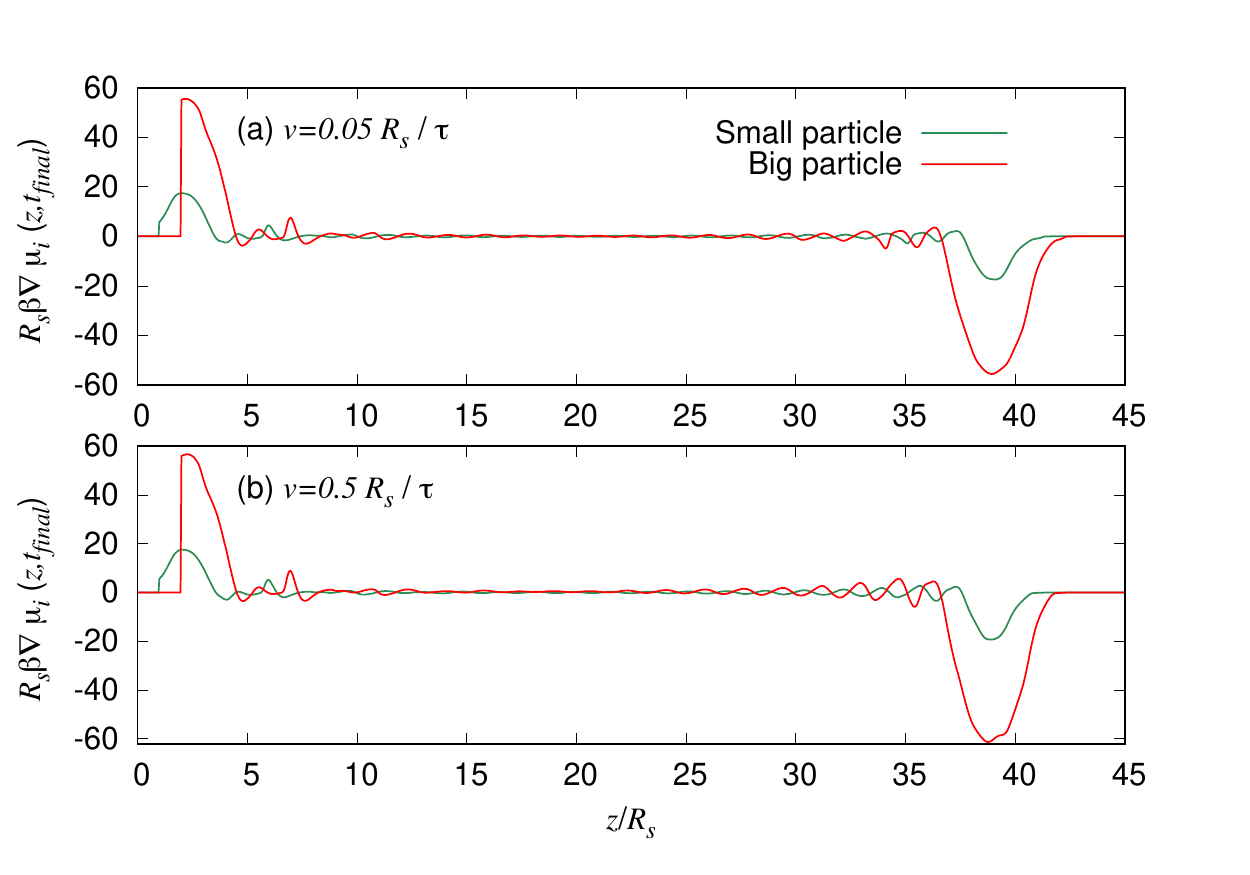}
	\caption{\label{fig:grad_chem} The gradient of the chemical potentials at the final state at $t=t_{final}$, corresponding to the results in Fig.~\ref{fig:different_speed}. These are for different interface descending velocities: (a) $v=0.05R_s/\tau$ and (b) $v=0.5R_s/\tau$.}		
\end{figure}

Further insight into the nature of the final states formed by the film drying process can be obtained from examining plots of the gradient of the local non-equilibrium chemical potentials, $\nabla\mu_i(\boldsymbol{r},t=t_{final})$. Recall from Eq.~\eqref{equ5} that these are the thermodynamic forces determining the time evolution of the system. In Fig.~\ref{fig:grad_chem} we display plots of these corresponding to the density profiles at $t=t_{final}$ shown in Fig.~\ref{fig:different_speed}. We see that in the bulk, away from the interfaces between $10\lesssim z/R_s\lesssim30$, that $\nabla\mu_i\approx 0$. This means that in this region, the particles are close to equilibrium. However, at the two interfaces, $\nabla\mu_i$ is very far from zero in value, indicating that in this region, the system is far from equilibrium. The size of the maxima and minima in $\nabla\mu_i$ at the ends of the system are proportional to forces at the interfaces, which are compressing the system. Interestingly, the two sets of curves in Fig.~\ref{fig:grad_chem} are rather similar, despite being for two very different values of the evaporation rate $v$. The similarity in these cases shows that our definition of the final time $t_{final}$ as the moment just before the failure of the algorithm we use to integrate forward in time the DDFT equations, also corresponds to the point in time when the force required to compress the system even further become large. Thus, $t_{final}$ is indeed roughly the time when the system becoming jammed.

\subsubsection{Influence of varying volume fraction}

\begin{figure*}[t]
	\includegraphics[scale=0.6]{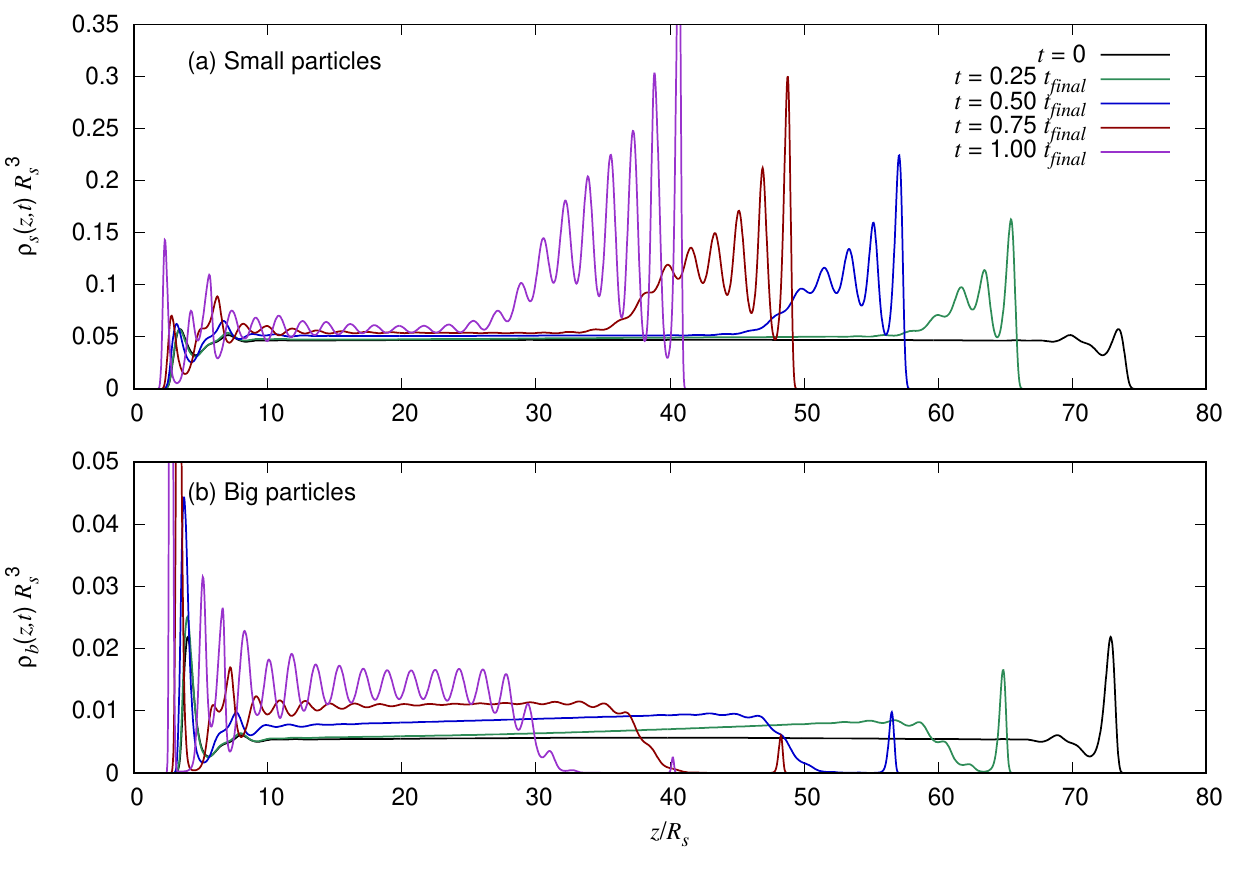}
	\includegraphics[scale=0.6]{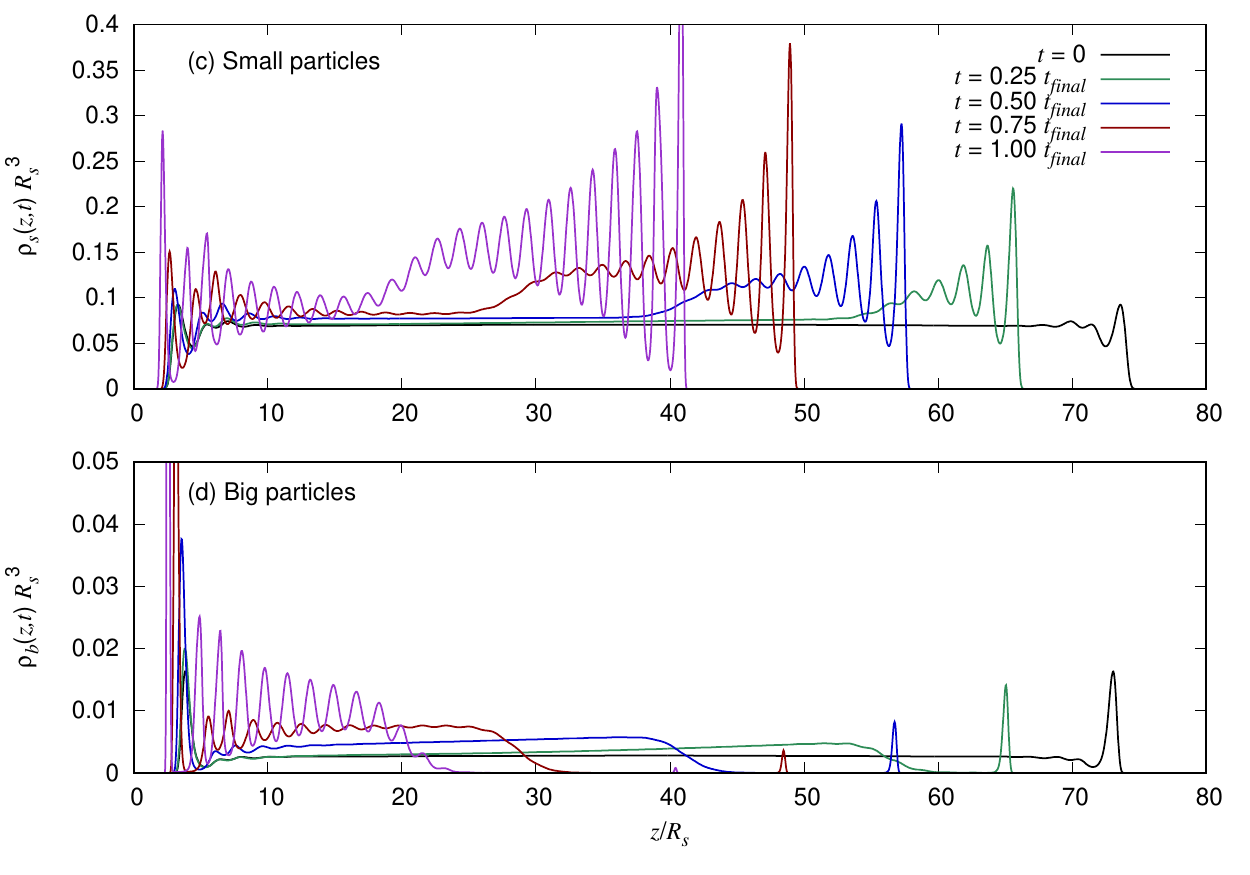}		
	\caption{\label{fig:different_volume} Time sequences of density profiles, for systems with an evaporation rate $v=0.5R_s/\tau$ (corresponding to $\mathrm{Pe}_s\approx40$ and $\mathrm{Pe}_b\approx80$), which leads to $t_{final}=68\tau$, and particle size ratio $\alpha=2$. In panels (a) and (b) on the left are the profiles for the small and big particles respectively, for the case when the initial average volume fractions are $\varphi_s = 0.2$ and $\varphi_b = 0.2$. On the right, in panels (c) and (d), are the results for $\varphi_s = 0.3$ and $\varphi_b = 0.1$. Corresponding results for the case when $\varphi_s = 0.1$ and $\varphi_b = 0.3$ are displayed in Fig.~\ref{fig:different_speed}(a)--(b). Some of the highest peaks in these density profiles are not displayed in order to show more details of the profiles away from the peaks. }  	
\end{figure*}

To illustrate the influence of varying the relative average volume fractions of the two species, in Fig.~\ref{fig:different_volume} we show density profiles varying over time for cases with $v = 0.5R_s/\tau$ and total initial volume fraction $\varphi\equiv\varphi_s+\varphi_b=0.4$, but with varying proportions of the big and the small particles. These results should be compared with the results in Fig.~\ref{fig:different_speed}(a)--(b), which are for the same values of $\varphi$ and $v$, but different concentrations, $\varphi_{s}=0.1$ and $\varphi_{b}=0.3$. The profiles in Fig.~\ref{fig:different_volume} are for an increased proportion of the small particles, namely $\varphi_{s}=0.2$ and $\varphi_b=0.2$ for the case in panels (a) and (b), and $\varphi_s = 0.3$ and $\varphi_b = 0.1$ for the case in (c) and (d). These show that the increased concentration of the small particles leads to an enhancement in the size segregation during drying. We move from a situation were the big particles are present throughout almost all of the final film, with only a narrow region a few particle diameters in thickness near the top of the film where the density of the big particles is low [see Fig.~\ref{fig:different_speed}(a)--(b)], to structures where big particles are completely depleted from a region of thickness about $8R_s$ in the case in Fig.~\ref{fig:different_volume}(a)--(b) and of thickness at least $15R_s$ for the case in Fig.~\ref{fig:different_volume}(c)--(d). This observation is in line with previous works which show that an increased number of small particles enhances the size segregation \cite{15zhou2017cross}. Note that the results in Fig.~\ref{fig:different_volume}(c)--(d) contradict the argument in Ref.~\cite{sear2018stratification}, where it was assumed that the small particles diffuse as an ideal-gas till they arrest when the local packing fraction reaches $\varphi_\textrm{jam}$ and that there is a maximum difusophoretic velocity of the big particles in front of a jammed layer of small particles, which is equal to $\frac{9}{4}v\varphi_\textrm{jam}$. This line of reasoning leads to the prediction that there is no segregation when $\varphi_s\gtrsim0.2$ \cite{sear2018stratification}. Our results in Fig.~\ref{fig:different_volume}(c)--(d) are above this threshold and we do still see segregation, likely stemming from the fact that our model does not incorporate the hydrodynamic interactions, albeit it goes well beyond the ideal-gas in treating the influence of the particle interactions. We discuss the implications of not modelling the solvent explicitly in a later section.

\subsubsection{Influence of the particle size ratio}

\begin{figure*}[t]
 	\includegraphics[scale=0.6]{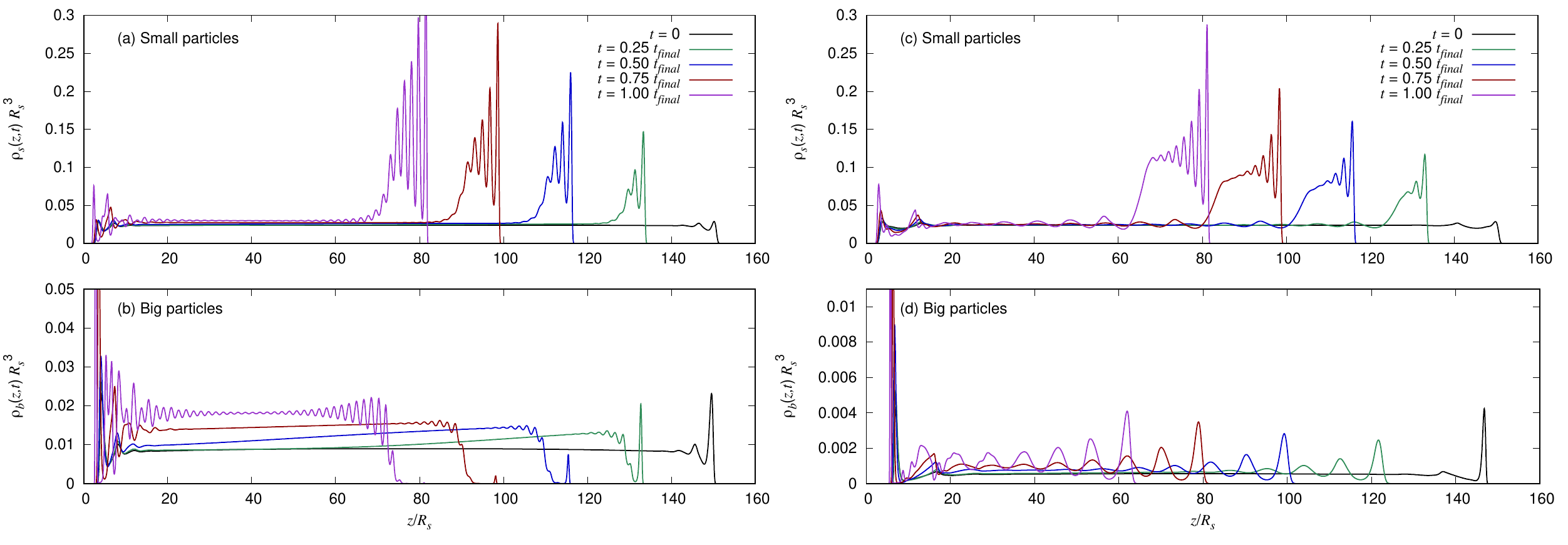}
\caption{\label{fig:different_size_ratio} 
Time sequences of density profiles, for systems with an evaporation rate $v=0.5R_s/\tau$, initial volume fractions $\varphi_s = 0.1$, $\varphi_b = 0.3$ and initial film thickness $L_0=154R_s$. These lead to $t_{final}=140\tau$. In panels (a) and (b) on the left, the size ratio $\alpha=2$; i.e.\ these are very similar to the system in Fig.~\ref{fig:different_speed}(a)--(b), except here the initial film is roughly double the thickness. This correspond to $\mathrm{Pe}_s\approx80$ and $\mathrm{Pe}_b\approx160$. On the right, in panels (c) and (d), are corresponding results for a system with much bigger size ratio, $\alpha=5$ ($\mathrm{Pe}_b\approx400$). In both cases a strong particle segregation is observed during drying. Some of the highest peaks in these density profiles are not displayed in order to show more details of the profiles away from the peaks. }
\end{figure*}

In Fig.~\ref{fig:different_size_ratio} we display results for the time evolution of the density profiles for systems with two different size ratios, $\alpha=2$ and $\alpha=5$. The interface velocity $v = 0.5R_s/\tau$, the initial film thickness $L_0=154R_s$ and the initial concentrations $\varphi_{s}=0.1$ and $\varphi_{b}=0.3$. Note that these systems have a film thickness that is initially roughly twice the thickness of the results displayed previously, so for the small particles this situation corresponds to $\mathrm{Pe}_s\approx80$ (since doubling $L_0$ also doubles the P\'eclet number). For the large particles, when $\alpha=2$ this gives $\mathrm{Pe}_b\approx160$ and when $\alpha=5$ gives $\mathrm{Pe}_b\approx400$. The larger $L_0$ allows the system to evolve for longer and so for a thicker layer of the small particles to develop at the upper interface, as can be seen by comparing the system with $\alpha=2$ in Fig.~\ref{fig:different_speed}(a)--(b) but having a thinner initial film. This trend is to be expected, since a thicker film takes longer to dry, giving more time for the particles to rearrange and segregate by size before jamming occurs. Increasing the size ratio to $\alpha=5$ increases the thickness of the small particle layer at the top interface [Fig.~\ref{fig:different_size_ratio}(c)--(d)]. Fortini et al.\ argue that it is the particle size that determines how fast the big particle population moves down in the concentration gradient of the small particles and therefore a larger size ratio results in a stronger size segregation \cite{12fortini2016dynamic}. However, note that when $R_b \gg R_s$ the drift speed of the particles is expected to be independent of the large particle size.\cite{anderson1989colloid} Therefore, it should be borne in mind that our implicit solvent model likely overestimates the degree of segregation as the size ratio is increased.

\subsection{Dilute systems}\label{subsec:dillute}

In the previous section we have discussed the influence of the evaporation rate, relative concentration and particle size ratio on the stratification for relatively dense systems, showing that higher evaporation rate, higher size ratio and higher initial volume fraction of small particles can enhance the emergence of the small-on-top structure. We now present results for dilute systems and show that stratification can occur in these also. We present DDFT results for cases with initial volume fractions $\phi_s=\phi_b=0.05$ and particle size ratio $\alpha=2$.
 
\subsubsection{Effect of evaporation rate}

\begin{figure*}[t!]
	\includegraphics[scale=0.6]{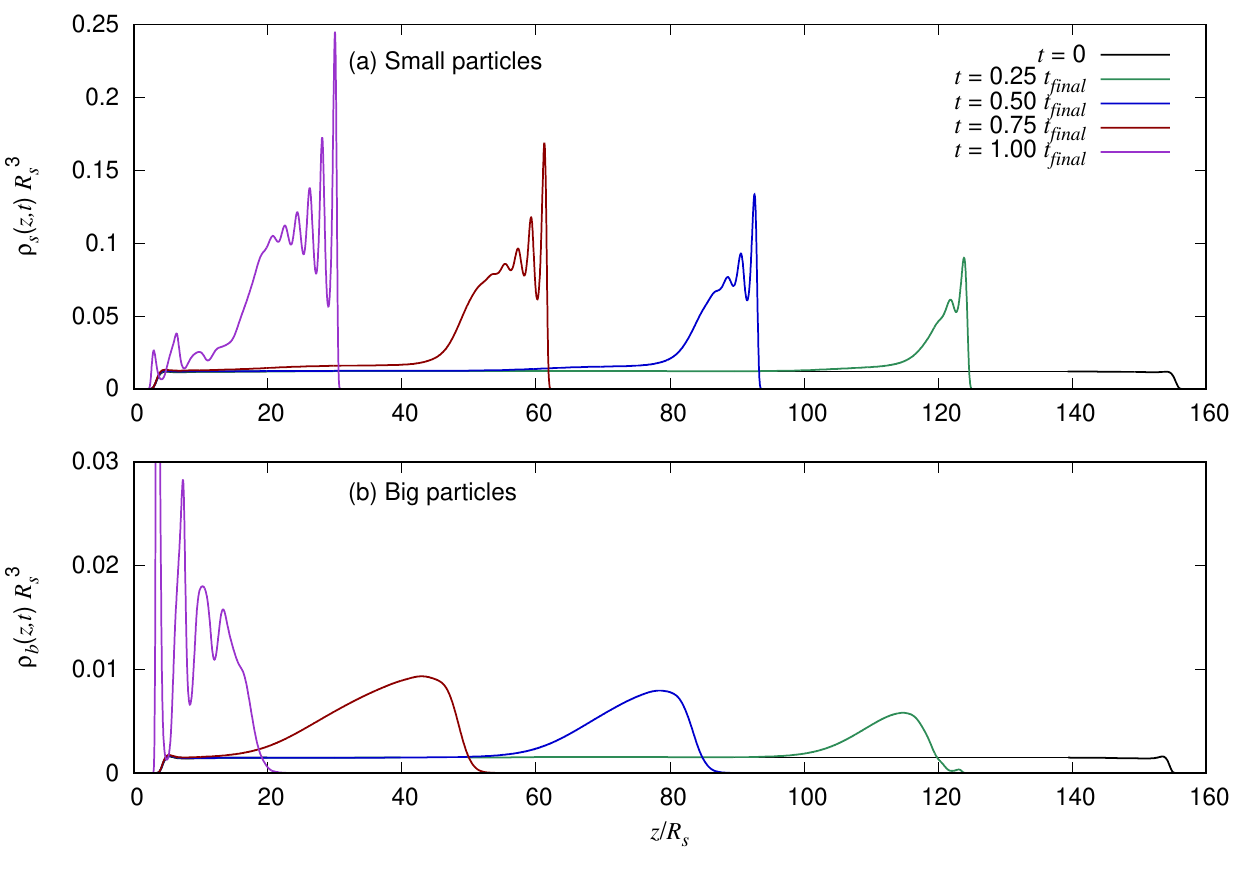}
	\includegraphics[scale=0.6]{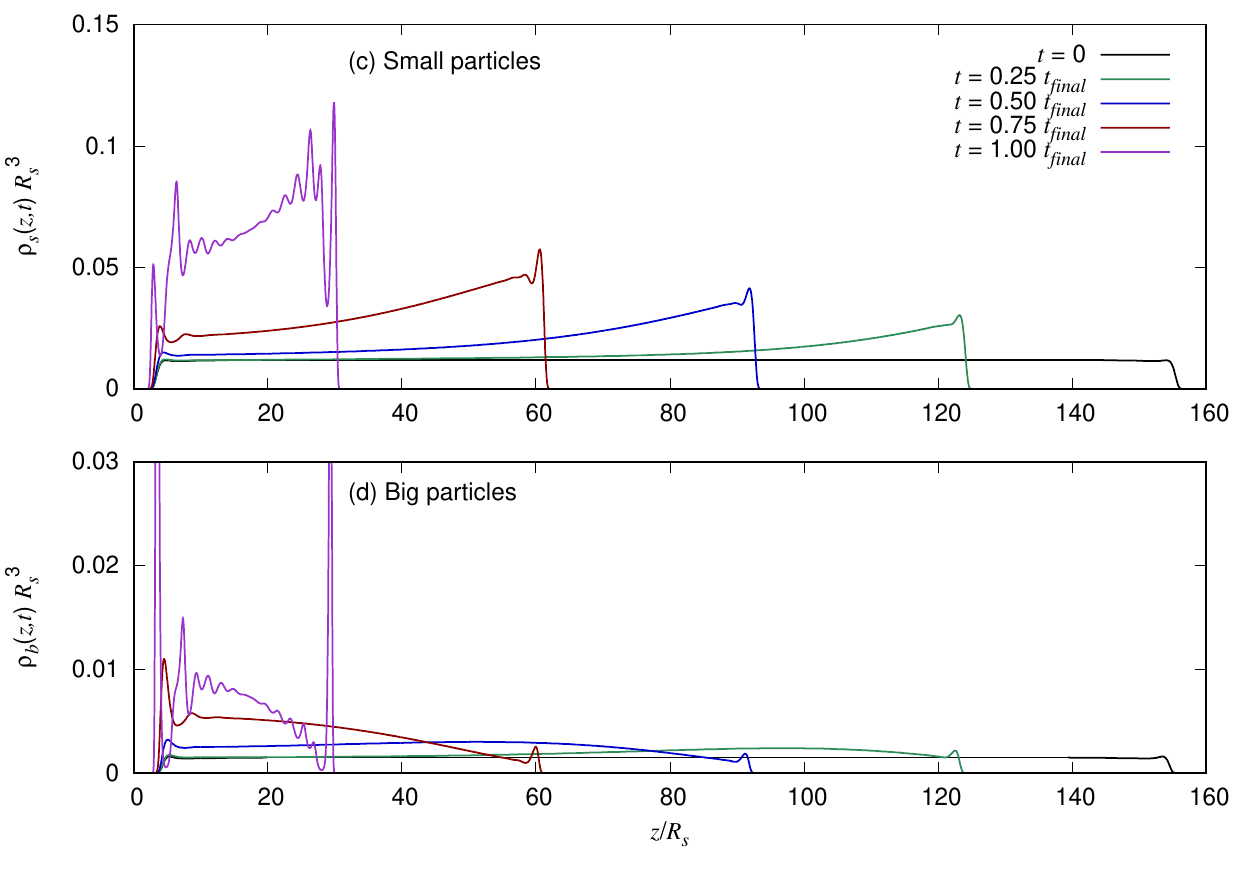}
	\caption{\label{fig:different_speed_dilute}
	Time sequences of density profiles for initially dilute systems having initial volume fractions $\varphi_s=\varphi_b = 0.05$, size ratio $\alpha=2$ and initial film thickness $L_0=154R_s$. In panels (a) and (b) on the left, the evaporation rate $v=0.5R_s/\tau$ (corresponding to $\mathrm{Pe}_s\approx80$ and $\mathrm{Pe}_b\approx160$), which leads to $t_{final}=255\tau$. In panels (c) and (d) on the right, $v=0.05R_s/\tau$ ($\mathrm{Pe}_s\approx8$ and $\mathrm{Pe}_b\approx16$), which leads to $t_{final}=2550\tau$. Some of the highest peaks in these density profiles are not displayed in order to show more details of the profiles away from the peaks. }  	
\end{figure*}

\begin{figure}[t]
	\includegraphics[scale=0.95]{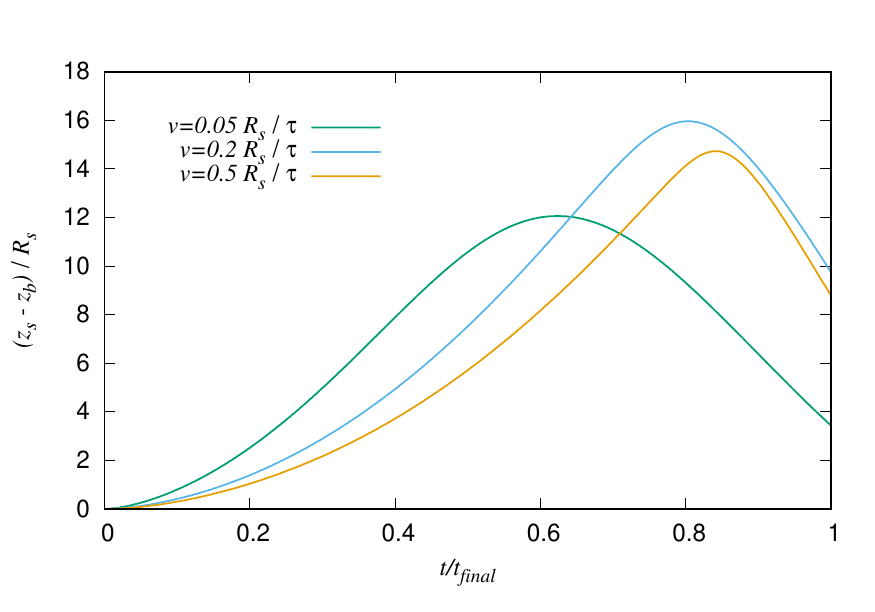}
	\caption{\label{fig:diff_center_of_mass_dilute} The time evolution of the difference between the positions of the centres of mass of the two species $(z_s-z_b)$, for various different interface descending velocities $v$, as given in the key. The size ratio of the particles $\alpha=2$ and the initial average volume fractions $\varphi_s = 0.05$ and $\varphi_b = 0.05$. For increasing $v$, as given in the key, corresponding to $\mathrm{Pe}_s\approx8$, 32 and 80 and $\mathrm{Pe}_b\approx16$, 64 and 160, and the final times are $t_{final}/\tau=2550$, $637$ and $255$.}		
\end{figure}

In Fig.~\ref{fig:different_speed_dilute}(a)--(b) are results for a case with $v=0.5R_s/\tau$ and $L_0=154R_s$. This is in a regime where the arguments of Trueman et al.~\cite{14trueman2012auto} would suggest the formation of big-on-top stratified structures, due to the big particles not being able to diffuse fast enough to escape the descending upper interface. This is not exactly what we observe, because right at the upper interface we see a build-up of the small particles. Nevertheless, at intermediate times, we do see a build-up of the big particles a short distance in-front of the descending upper interface for the reasons identified by Trueman et al. This can be seen from the `hump' in the big particle density profiles in Fig.~\ref{fig:different_speed_dilute}(b) for times $0.25\lesssim t/t_{final}\lesssim 0.8$. The region of the hump correspond to a layer where there is a build-up of the big particles. However, in the final stages of the dynamics, when the system becomes much denser, this region with enhanced density of the big particles disappears. The final structure consists of a layer at the top almost entirely of small particles, below which is a layer dominated by the big particles, but still containing some small particles.

In Fig.~\ref{fig:different_speed_dilute}(c)--(d) are displayed the corresponding results for a case with a much slower interface descending velocity, $v=0.05R_s/\tau$. At this lower velocity, the particles have enough time to move away from the interface by diffusion, preventing any sizeable concentration gradients forming or associated stratification \cite{12fortini2016dynamic}. This is in contrast with the case in Fig.~\ref{fig:different_speed_dilute}(a)--(b), which is for a much higher interface velocity. That said, a gradient in the density distributions of the particles through the film does gradually develop over time, together with and associated very weak segregation.

In Fig.~\ref{fig:diff_center_of_mass_dilute} we display plots of the difference between the centre of mass positions $(z_s-z_b)$ over time, for three different values of the interface velocity $v$. Two of these results correspond to the density profiles in Fig.~\ref{fig:different_speed_dilute}. Interestingly, all of the curves in Fig.~\ref{fig:diff_center_of_mass_dilute} feature a maximum, indicating the existence of a maximally segregated state of the system at an intermediate time, where the peak occurs. The location of the maximum moves to latter times with increasing evaporation rate $v$, and the time at which the maximum occurs approaches $t_{final}$ for large values of $v$. The value of the maximum (i.e.\ the degree of segregation) also increases with increasing $v$. The fact that $(z_s-z_b)$ decreases again in the later stages for the smaller $v$ cases is associated with the fact that these systems are only weakly segregated in the final state.

\subsubsection{Larger size ratio}

\begin{figure}[t]

	\includegraphics[scale=0.67]{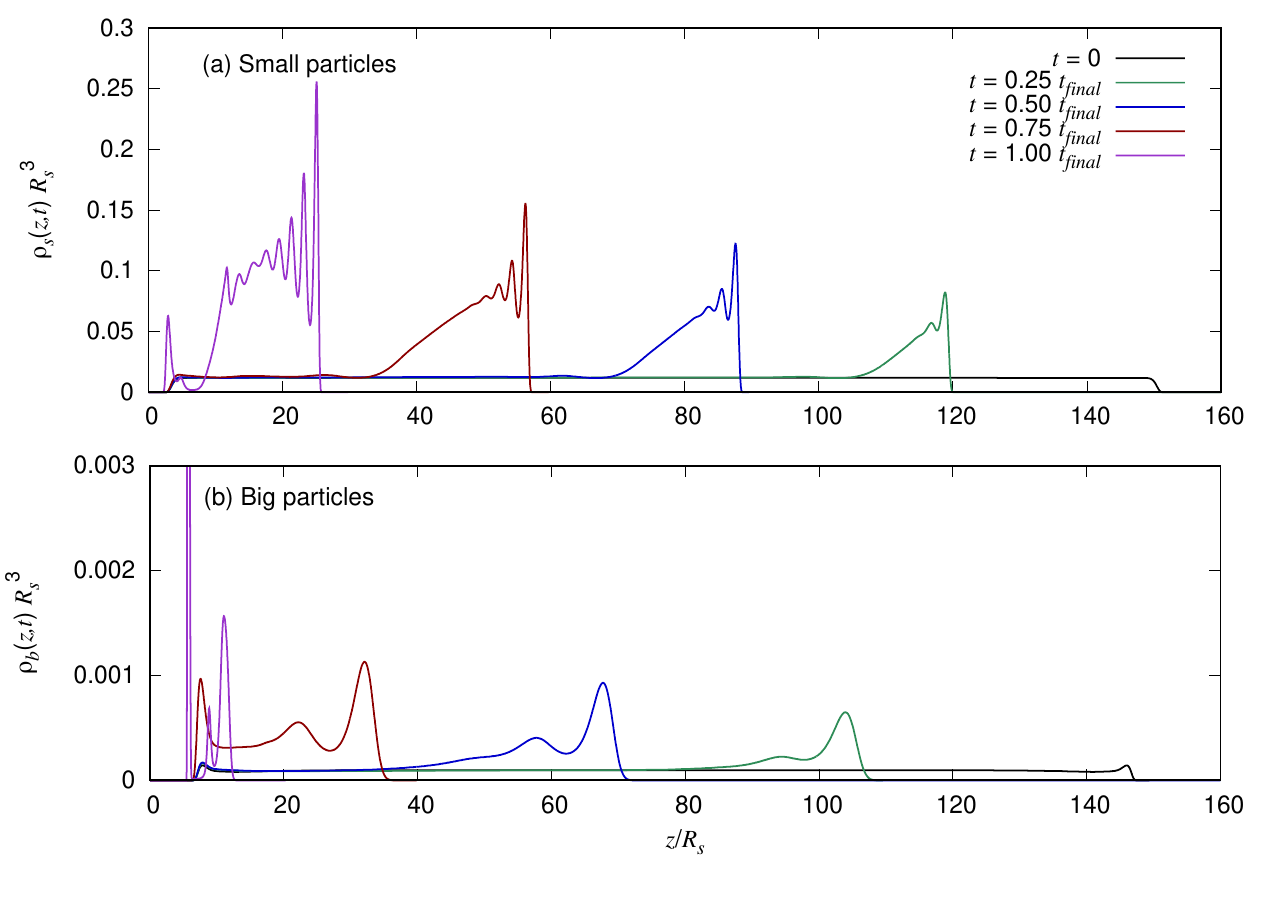}
	
	\caption{\label{fig:different_size_ratio_dilute} 
		Time sequence of density profiles for an initially dilute system with $\varphi_s = \varphi_b = 0.05$, size ratio $\alpha=5$ and evaporation rate $v=0.5R_s/\tau$ (which corresponds to $\mathrm{Pe}_s\approx80$ and $\mathrm{Pe}_b\approx400$). The highest peaks in $\rho_b(z,t_{final})$ are not displayed in order to show more details of the profiles away from the peaks.}  	
\end{figure}

Figure~\ref{fig:different_size_ratio_dilute} shows a time sequence of density profiles for a case with the larger size ratio $\alpha=5$, while all other parameters are the same as for Fig.~\ref{fig:different_speed_dilute}(a)--(b). In this case, the thickness of the small-particle-only region at the top is significantly increased to about $20R_s$ at $t=0.75t_{final}$, but in the final stages shrinks to about $15R_s$, as the system is compressed to form a dense system. This observation is in accordance with the results of previous studies reporting enhanced size segregation in dilute systems for higher size ratios in both experiments and simulations \cite{makepeace2017stratification}. Owing to the fact that there is a thick layer of small particles on top and also a density peak in the small particle profile right at the lower surface, with the big particles in-between, the final structure in Fig.~\ref{fig:different_size_ratio_dilute} has some similarities to the ``small-large-small sandwich'' structure reported by Liu et al.~\cite{liu2019sandwich}. However, they observed this structure in the size ratio range $1.2 \lesssim \alpha \lesssim 2.8$, while a different large-small-large structure developed for $\alpha > 2.8$.

\subsection{Considerations on treating the solvent implicitly}

In the limit where the density of the big particles is small and where the density distribution of the small particles is slowly varying, we may calculate the diffusiophoretic drift velocity $\boldsymbol{U}_{b}=\boldsymbol{J}_b/\rho_b$ predicted by the DDFT, where the current $\boldsymbol{J}_b$ is given in Eq.~\eqref{equ5}. A useful result in the low-density limit when the densities are constants is\cite{roth2010fundamental}
\begin{equation}
\mu_b=k_BT\left[\ln\Lambda_b^3\rho_b-\rho_b\int d\boldsymbol{r} f_{bb}(r)-\rho_s\int d\boldsymbol{r} f_{bs}(r)\right] + O(\rho_i^2),
\label{eq:mu_b}
\end{equation}
where $f_{ij}(r)$ are Mayer functions, which for hard-spheres we have $f_{ij}(r)=-1$ for $r<R_i+R_j$ and $f_{ij}(r)=0$ otherwise. This gives $\int d\boldsymbol{r} f_{ij}(r) = -\frac{4}{3}\pi(R_i+R_j)^3$. Then, if $\rho_b$ is constant and the density of the small particles is slowly varying, we can write $\nabla\mu_b\approx\frac{\partial \mu_b}{\partial \rho_s}\nabla\rho_s$, which together with Eq.~\eqref{eq:mu_b} gives
$\boldsymbol{U}_b=-\Gamma_bk_BT\frac{4}{3}\pi(R_b+R_s)^3\nabla\rho_s$. Or, using $\Gamma_b=D_b/k_BT$ together with the Stokes-Einstein relation in Eq.~\eqref{eq:Stokes-Einstein}, gives
\begin{equation}
\boldsymbol{U}_b=-\frac{2}{9\eta}k_BTR_b^2\left(1+\frac{1}{\alpha}\right)^3\nabla\rho_s.
\label{eq:DDFT_estimate}
\end{equation}
When the size ratio $\alpha$ is large, this amounts to balancing an osmotic buoyancy force $\sim R_b^3 k_BT \nabla\rho_s$ against Stokes’ drag $\sim6\pi\eta R_b \boldsymbol{U}_b$, giving a diffusiophoretic drift velocity $\boldsymbol{U}_b \sim (R_b^2 k_BT / \eta) \nabla\rho_s$, which is strongly dependent on the size of the big particles. However, as explained by J\"ulicher and Prost,\cite{julicher2009comment} Brady,\cite{brady2011particle} Sear and Warren,\cite{sear2017diffusiophoresis} and most recently by Marbach et al.~\cite{marbach2020local}, the osmotic buoyancy force is almost totally cancelled by the pressure arising from solvent backflow, leaving only a residual effect. This difference is particularly relevant when $R_b \gg R_s$, i.e.\ when $\alpha\gg1$. In this case, the actual drift velocity is likely independent of the size of the big particles,\cite{anderson1989colloid, marbach2020local} corresponding to an effective slip in a boundary layer on the surface of the large colloids of thickness $\sim R_s$, giving $\boldsymbol{U}_b \sim (R_s^2 k_BT / \eta) \nabla\rho_s$. This diffusiophoretic drift velocity is a factor $\alpha^2$ smaller than the DDFT estimate. Thus, on the basis of these low-density limit arguments, as mentioned in the introduction, explicitly treating the hydrodynamics will not change the direction of the observed stratification, but will likely make the effect less marked than we observe in the present implicit-solvent model treatment. That said, in the later stages of the drying process, the local particle densities become large, and we expect the hydrodynamic forces to be `screened' to some extent in such a crowded environment, where the above arguments no longer apply.

\section{\label{sec:4}Concluding remarks}

In this paper we have studied the drying of binary colloidal dispersions using DDFT in conjunction with an accurate FMT approximation for the Helmholtz free energy functional \cite{roth2010fundamental}. Our model treats the colloids as neutral hard-spheres and we assume that the density profiles only vary in the direction perpendicular to the planar surface, so that the DDFT equations we have solved are one-dimensional. We have treated the influence of the solvent evaporation by modelling the upper solvent-air interface as a planar surface that descends with constant velocity (evaporation rate) $v$, i.e.\ as a time-dependent external potential that moves with constant velocity toward the stationary lower surface.

Three parameters that are known to play a key role in the assembly of drying binary colloidal dispersions have have been examined here: evaporation rate, initial volume fraction and size ratio. In the dense systems studied here, we have observed stratification which, in agreement with the experiments of Makepeace et al.\ \cite{makepeace2017stratification}, is enhanced when the the particle size ratio is larger and there is a higher initial volume fraction of the small particles -- see e.g.\ Figs.~\ref{fig:different_volume} and \ref{fig:different_size_ratio}. This observed stratification is in-line with the arguments of Fortini et al.\ \cite{12fortini2016dynamic}, namely that bigger particles can move down faster due to the concentration gradient.

We have also examined how the evaporation rate affects the final film structure -- see e.g.\  Fig.~\ref{fig:different_speed}. The results show that as the evaporation rate is increased, the particle segregation increases and so at the higher evaporation rates the system exhibits stratification. The degree of segregation can be quantified by calculating the difference between the distances of the centres of mass of the two species of particles from the lower surface during the drying process -- see e.g.\ Fig.~\ref{fig:diff_centre_of_mass}. In dilute systems, the effect of evaporation rate can also be significant, including leading to transient states with an enhancement in the concentration of the big particles near the top, due to their slow diffusion rates and being unable to escape from the advancing interface. However, once such systems enter the dense-suspension regime near the end of the drying process, we find that the system either remains mixed or that the small particles end up in a layer on the top of the system. Our results for dilute systems are in line with the state diagram proposed by Zhou et al.~\cite{15zhou2017cross}.

Our work further highlights the importance of the size ratio, initial volume fraction and evaporation rate on controlling the stratification of films drying from binary colloidal mixtures. Note that in our model the hydrodynamic interactions between the colloids have been neglected, which does not change the direction of the stratification but may lead to an overestimate in the degree to which it is predicted to occur.\cite{16tang2018stratification, sear2017diffusiophoresis} Hydrodynamic interactions can be included in DDFT \cite{rex2008dynamical, goddard2016dynamical, goddard2012general}, but we leave this to future work. Similarly, it would be instructive to apply power functional theory,\cite{schmidt2013power, fortini2014superadiabatic, de2019custom, heras2020flow} to incorporate into the modelling the superadiabiatic forces that our DDFT neglects, in order to determine the influence of these on the observed stratification. It would also in the future be interesting to consider variations in the density distribution of the particles parallel to the surface, because especially if there are also attractive interactions between the colloids, one should expect the particle density distributions to vary in both the perpendicular and parallel directions to the surface \cite{malijevsky2013sedimentation}.

\section{Acknowledgements}

We are grateful to the EPSRC for the PhD studentship funding held by B.H.

\providecommand{\noopsort}[1]{}\providecommand{\singleletter}[1]{#1}%
\providecommand{\latin}[1]{#1}
\makeatletter
\providecommand{\doi}
  {\begingroup\let\do\@makeother\dospecials
  \catcode`\{=1 \catcode`\}=2 \doi@aux}
\providecommand{\doi@aux}[1]{\endgroup\texttt{#1}}
\makeatother
\providecommand*\mcitethebibliography{\thebibliography}
\csname @ifundefined\endcsname{endmcitethebibliography}
  {\let\endmcitethebibliography\endthebibliography}{}


\end{document}